\documentclass[reprint,amsmath,amssymb,aps,pre,showpacs,eqsecnum,]{revtex4-1}

\usepackage{graphicx}
\usepackage{amsbsy}
\usepackage{amsfonts}
\usepackage{float}
\usepackage{color}
\usepackage{bm}% bold math

\newcommand{\bq}{\begin{eqnarray}}
\newcommand{\eq}{\end{eqnarray}}
\newcommand{\bqn}{\begin{eqnarray*}}
\newcommand{\eqn}{\end{eqnarray*}}

\newcommand\beq{\begin{equation}}
\newcommand\eeq{\end{equation}}
\newcommand\beqa{\begin{eqnarray}}
\newcommand\eeqa{\end{eqnarray}}
\newcommand{\nn}{\nonumber\\}
\newcommand{\zero}{{(0)}}
\newcommand{\one}{{(1)}}

\newcommand{\aB}{\text{RFA}_+}
\newcommand{\s}{\tau}
\newcommand{\ellw}{w}

\begin{document}
%%%%%%%%%%%%%%%%%%%%%%%%%%%%%%%%%%%%%%%%%%%%%%%%%%%%%%%%%%%%%%%%%%%%%%%%%%%%%%
%%%%%%%%%%%%%%%%%%%%%%%%%%%%%%%%%%%%%%%%%%%%%%%%%%%%%%%%%%%%%%%%%%%%%%%%%%%%%%
%%%%%%%%%%%%%%%%%%%%%%%%%%%%%%%%%%%%%%%%%%%%%%%%%%%%%%%%%%%%%%%%%%%%%%%%%%%%%%
\title{Multicomponent fluid of nonadditive hard spheres near a wall}

\author{Riccardo Fantoni}
\email{rfantoni@ts.infn.it}
\affiliation{Dipartimento di Scienze dei Materiali e Nanosistemi,
  Universit\`a Ca' Foscari Venezia, Calle Larga S. Marta DD2137, I-30123
  Venezia, Italy}

\author{Andr\'es Santos}
\email{andres@unex.es}
\homepage{http://www.unex.es/eweb/fisteor/andres}
\affiliation{Departamento de F\'isica, Universidad de Extremadura,
  E-06071 Badajoz, Spain}

\date{\today}

\begin{abstract}
A recently proposed rational-function approximation [Phys.\ Rev.\ E \textbf{84}, 041201 (2011)] for the structural properties of nonadditive hard spheres is applied to evaluate analytically (in Laplace space)  the local density profiles of multicomponent nonadditive hard-sphere mixtures near a planar nonadditive  hard wall. The theory is assessed by comparison with $NVT$ Monte Carlo simulations of binary mixtures with a size ratio 1:3 in three possible scenarios:  a  mixture with either positive or negative nonadditivity near an additive wall, an additive mixture with a {nonadditive wall}, and a nonadditive mixture with a nonadditive wall. It is observed that, while the theory tends to underestimate the local densities at contact (especially in the case of the big spheres) it captures very well the initial decay of the densities with increasing separation from the wall and the subsequent oscillations.
\end{abstract}

%\keywords{Wall, Nonadditive Hard Sphere, Percus Yevick Approximation,  Rational Function Approximation, Monte Carlo Simulation.}

\maketitle
%%%%%%%%%%%%%%%%%%%%%%%%%%%%%%%%%%%%%%%%%%%%%%%%%%%%%%%%%%%%%%%%%%%%%%%%%%%%%%
\section{Introduction}
%%%%%%%%%%%%%%%%%%%%%%%%%%%%%%%%%%%%%%%%%%%%%%%%%%%%%%%%%%%%%%%%%%%%%%%%%%%%%%
\label{sec:introduction}

The study of mixtures near a fluid-solid interface is important for the understanding of wetting
and adsorption phenomena where competition among  different components may occur. A simplified
physical picture of adsorption may be obtained at a microscopic level
if one considers the solid surface as a planar smooth hard wall
confining the particles of the mixture. Thereby, one can describe
the expected oscillations of the (partial) local particle densities in the
neighborhood of the wall with an abundance of particles right at
contact and a depletion nearby. Whereas confined fluid mixtures of
additive hard spheres (AHS) have been widely studied  within integral equation
theories
\cite{HAB76,H78,PH84,PH85,HCD94,DAS97,ODHQ97,NHSC98},
Monte Carlo simulations
\cite{NHSC98,TMSG89,SH78,DH93,RNG05,MYSH07}, and density-functional theories \cite{TMSG89,PG97,P99,RD00,ZR00,Z01,CG01,PG02a,PG02b,PG02c,PG03,MYSH07},
 much less is known in the case of nonadditive hard spheres (NAHS)
\cite{DVA03,PSP05,JDOL08,HS11b}.

In a recent paper \cite{FS11},   NAHS mixtures were studied through the so-called rational-function
approximation (RFA) technique \cite{YHS96,HYS08}, which amounts to choosing
simple (rational-function) expressions for the Laplace space
representation of the radial distribution functions of the  theory of liquids \cite{HM06,FP03}. This
allowed us to determine a nonperturbative,
fully analytical (in Laplace space) approximation. When the
nonadditivity is set to zero, the approximation reduces to the Percus--Yevick
(PY) approximation for an AHS mixture.

The purpose of the present work is to use the RFA scheme devised in
Ref.\ \cite{FS11} to determine the structural properties
of an $n$-component NAHS fluid  near a hard wall interacting either additively or nonadditively with the particles of the fluid mixture. A realization
of the problem is obtained from a $(n+1)$-component NAHS mixture, where
one of the species, species $0$, is taken to have  a vanishing concentration and an infinite diameter. A similar approach was employed by Malijevsky {\em et al.} \cite{MYSH07} to determine through
the RFA the structural properties of a multicomponent AHS fluid
near an additive hard wall.
In the present case, however, not only the particle-particle interaction may be nonadditive (i.e., the closest distance between the centers of two spheres of species $i$ and $j$ is in general different from the arithmetic mean of the respective diameters), but also the particle-wall may be nonadditive as well. The latter possibility means that the closest distance from the planar wall to the center of a sphere may be different from the radius of the sphere. A similar problem has recently been considered by Gonz\'alez \emph{et al.} \cite{GWRV11},
where strong size selectivity is observed in a binary AHS mixture confined in a narrow cylindrical
pore such that each species of the mixture sees a different cylinder radius.

We will compare our approximation results for the local density of particles
at a  distance $z$ from the wall with exact canonical (fixed
number of particles $N$, volume $V$, and temperature $T$) Monte Carlo
(MC) simulation results for binary mixtures. In the simulation it is
necessary to use two hard walls on the opposite far square faces of a
parallelepiped simulation box with rectangular lateral faces
and to choose the two walls far enough so that bulk properties of the
fluid can be extracted by looking at the center of the box.

{The agreement between theory and simulations is quite satisfactory. It is worse at contact (similarly to what happens with the PY theory in the additive case \cite{MYSH07}) but it rapidly improves as the distance from the wall increases, so that the first minimum (depletion region) and the subsequent oscillations are well predicted by our analytical approach. To the best of our knowledge, our results constitute the first proposal for an
analytical expression (in Laplace space) for the density profiles of a
NAHS mixture confined by a (nonadditive or additive) hard wall. As such, the theory is expected to be useful to the experimentalist who needs easy formulas to determine
profiles to compare with  experimental data, thus bypassing the need of
numerical experiments.}

{The paper is organized as follows. In Sec.\ \ref{sec:model} we describe
the model of the confined fluid we are going to study.
The RFA used to extract the structural
properties of the fluid is presented in Sec.\ \ref{sec:rfa}, where some details of the wall limit are
given in  the Appendix. In Sec.\ \ref{sec:mc} we describe some details of the $NVT$ MC simulation method we employed for confined binary mixtures. The results for
the structural properties are presented in Sec.\ \ref{sec:results},
where the RFA and our own MC simulation are compared. Finally,
Sec.\ \ref{sec:conclusions} is left for concluding remarks.}

%%%%%%%%%%%%%%%%%%%%%%%%%%%%%%%%%%%%%%%%%%%%%%%%%%%%%%%%%%%%%%%%%%%%%%%%%%%%%%%
\section{The model}
%%%%%%%%%%%%%%%%%%%%%%%%%%%%%%%%%%%%%%%%%%%%%%%%%%%%%%%%%%%%%%%%%%%%%%%%%%%%%%%
\label{sec:model}

An $n$-component NAHS mixture  in the
$d$-dimensional Euclidean space is a fluid of $N_i$ particles of
species $i$ with $i=1,2,\ldots,n$, such that there are a total number
of particles $N=\sum_{i=1}^n N_i$ in a
volume $V$, and the pair potential between a particle of species $i$
and a particle of species $j$ separated by a distance $r$ is given by
\beq
\phi_{ij}(r)=\left\{\begin{array}{ll}
\infty, & r\leq\sigma_{ij}, \\
0 ,      & r>\sigma_{ij},
\end{array}\right.
\eeq
where $\sigma_{ii}=\sigma_i$ and
$\sigma_{ij}=\frac{1}{2}(\sigma_i+\sigma_j)(1+\Delta_{ij})$, so that
$\Delta_{ii}=0$ and $\Delta_{ij}=\Delta_{ji}>-1$. When
$\Delta_{ij}=0$ for every pair $i$-$j$ we recover the AHS system. In
the present paper we will only consider the NAHS system in its single fluid
phase.

Let ${\bar{\rho}}=N/V$ be the total number density of the mixture and {$\bar{x}_i=N_i/N$ be the mole fraction
of species $i$. These are spatially averaged quantities that can differ from local values in confined situations}.

The one-dimensional ($d=1$) NAHS fluid admits an exact analytical solution for the structural and
thermophysical properties in the thermodynamic limit $N\to\infty$ with ${\bar{\rho}}=N/V=\text{const}$
\cite{SZK53,LZ71,HC04,S07}. Moreover, the AHS fluid with
$d=\text{odd}$ allows for an
analytical solution of the PY approximate theory \cite{L64,YSH98,RS11,RS11b}. Such a solution in the case $d=1$ reduces to the exact solution particularized to the additive mixture.

Inspired by both  the exact solution for one-dimensional NAHS mixtures and the  PY solution for three-dimensional AHS mixtures, we have recently proposed an analytical approach for the three-dimensional NAHS system \cite{FS11}. As said in Sec.\ \ref{sec:introduction}, the aim  of the present paper is to use that approximation to determine the structural properties of a ternary
mixture where  {one of the species ($i=0$)} is subject to the {\em wall limit}:
${\bar{x}}_0\to 0$ and
$\sigma_0\to\infty$. Such a {ternary} mixture represents a binary mixture of {AHS ($\Delta_{12}= 0$) or}
NAHS ($\Delta_{12}\neq 0$) in the presence of a hard wall which, in addition, may interact {additively or} nonadditively with the fluid particles {(see Sec.\ \ref{walllimit})}.

%%%%%%%%%%%%%%%%%%%%%%%%%%%%%%%%%%%%%%%%%%%%%%%%%%%%%%%%%%%%%%%%%%%%%%%%%%%%%%
\section{Rational-function approximation}
%%%%%%%%%%%%%%%%%%%%%%%%%%%%%%%%%%%%%%%%%%%%%%%%%%%%%%%%%%%%%%%%%%%%%%%%%%%%%%
\label{sec:rfa}

\subsection{General scheme}

In Ref.\ \cite{FS11}, the following proposal for the
structural properties of an $n$-component NAHS fluid defined through the
Laplace transform $G_{ij}(s)$ of $rg_{ij}(r)$ was given:
\beq
G_{ij}(s)=s^{-2}\sum_{k=1}^n e^{-\sigma_{ik}s}L_{ik}(s)B_{kj}(s),
\label{Gij}
\eeq
with
\beq
\mathsf{B}^{-1}(s)= \mathsf{I}-\mathsf{A}(s),
\label{Bij}
\eeq
\beq
A_{ij}(s)= \frac{2\pi{\bar{\rho}}
{\bar{x}}_i}{s^3}\left[N_{ij}(s)e^{a_{ij}s}-L_{ij}(s)e^{-\sigma_{ij}s}\right],
\label{Qij}
\eeq
where $\mathsf{I}$ is the unit matrix,
\beq
{L}_{ij}(s)\equiv {L}_{ij}^\zero+{L}_{ij}^\one s,
\eeq
\beq
N_{ij}(s)\equiv L_{ij}^\zero\left(1-b_{ij} s+\frac{b_{ij}^2
  s^2}{2}\right)+L_{ij}^\one s\left(1-b_{ij} s\right),
\label{Nkj}
\eeq
\beq
b_{ij}\equiv
\sigma_{ij}+a_{ij}, \quad a_{ij}\equiv \frac{1}{2}(\sigma_i-\sigma_j).
\label{bij}
\eeq
Equations \eqref{Gij}--\eqref{Nkj} provide the explicit $s$-dependence of the Laplace transform $G_{ij}(s)$, but it still remains to obtain the two sets of parameters $L_{ij}^\zero$ and $L_{ij}^\one$. This is done by enforcing the physical requirements $\lim_{s\to 0}s^2 G_{ij}(s)=1$ and $\lim_{s\to 0}s^{-1}\left[s^2 G_{ij}(s)-1\right]=0$ \cite{FS11}. The result is
\beq
L_{ij}^\zero=S_j,\quad
L_{ij}^\one=T_j+\sigma_{ij}S_j,
\label{Lij0}
\eeq
where
\beq
S_j\equiv\frac{1-\pi {\bar{\rho}}\Psi_j}{\left(1-\pi
{\bar{\rho}}\Lambda_j\right)\left(1-\pi
{\bar{\rho}}\Psi_j\right)-\pi^2{\bar{\rho}}^2\mu_{j|2,0}\Omega_j},
\label{28a}
\eeq
\beq
T_j\equiv\frac{\pi{\bar{\rho}}\Omega_j}{\left(1-\pi {\bar{\rho}}\Lambda_j\right)\left(1-\pi
{\bar{\rho}}\Psi_j\right)-\pi^2{\bar{\rho}}^2\mu_{j|2,0}\Omega_j},
\label{29a}
\eeq
\beq
\Lambda_j\equiv \mu_{j|2,1}-\frac{1}{3}\mu_{j|3,0},
\label{30a}
\eeq
\beq
\Psi_j\equiv \frac{2}{3}\mu_{j|3,0}-\mu_{j|2,1},
\label{31}
\eeq
\beq
\Omega_j\equiv \mu_{j|3,1}-\mu_{j|2,2}-\frac{1}{4}\mu_{j|4,0},
\label{32a}
\eeq
and we have called
\beq
\mu_{j|p,q}\equiv \sum_{k=1}^n {\bar{x}}_k b_{kj}^p \sigma_{kj}^q.
\label{25}
\eeq

As discussed in Ref.\ \cite{FS11}, the inverse Laplace transform $\mathcal{L}^{-1}\left[G_{ij}(s)\right](r)$ may present a spurious behavior in the shell $\min(\sigma_{ij},\s_{ij})\leq r\leq \max(\sigma_{ij},\s_{ij})$, where $\s_{ij}$ is the minimum of the list of values $\sigma_{ik}-a_{kj}$ ($k=1,\ldots,n$) that are different from $\sigma_{ij}$. If $\sigma_{ik}-a_{kj}=\sigma_{ij}$ for all $k$, then $\s_{ij}=\sigma_{ij}$. The anomalous behavior of $\mathcal{L}^{-1}\left[G_{ij}(s)\right](r)$ for $\min(\sigma_{ij},\s_{ij})\leq r\leq \max(\sigma_{ij},\s_{ij})$ can be avoided with a series of corrections, the simplest one of which yields
\beqa
g_{ij}(r)&=&\Theta(r-\sigma_{ij})\left[\frac{\mathcal{L}^{-1}\left[G_{ij}(s)\right](r)}{r}\right.\nn
&&\left.+C_{ij}\Theta(\s_{ij}-r)\left(\frac{\s_{ij}}{r}-1\right)\right],
\label{54}
\eeqa
where
\beq
C_{ij}=
2\pi{\bar{\rho}} {\bar{x}}_{\kappa_{ij}} L_{i\kappa_{ij}}^\one\left(
L_{\kappa_{ij} j}^\one-S_j\frac{b_{\kappa_{ij} j}}{2}\right)b_{\kappa_{ij} j},
\label{gamma1}
\eeq
{$\kappa_{ij}$} being the index associated with $\s_{ij}$; i.e.,
$\s_{ij}=\sigma_{i\kappa_{ij}}-a_{\kappa_{ij} j}$.
The contact values are given by \cite{FS11}
\beq
g_{ij}(\sigma_{ij}^+)=\frac{L_{ij}^\one}{\sigma_{ij}}+C_{ij}\left(\frac{\s_{ij}}{\sigma_{ij}}-1\right).
\label{contact}
\eeq

The approximation \eqref{54} was referred to as $\aB^{(1)}$ in Ref.\ \cite{FS11}.
In the special case of AHS mixtures, one has $\sigma_{ik}-a_{kj}=\sigma_{ij}$, so that $\s_{ij}=\sigma_{ij}$ and $g_{ij}(r)=r^{-1}{\mathcal{L}^{-1}\left[G_{ij}(s)\right](r)}$ coincides with the PY solution \cite{L64,YSH98}.

\subsection{Wall limit}
\label{walllimit}
Now we assume that a single sphere of diameter $\sigma_0$ is introduced in the $n$-component fluid. This gives rise to an $(n+1)$-component fluid, where the extra species ($i=0$), being made of a single particle, has a vanishing concentration ${\bar{x}}_0=0$ in the thermodynamic limit $N\to\infty$. With this proviso, Eq.\ \eqref{Gij} can be easily extended to this $(n+1)$-component mixture.

According to Eq.\ \eqref{Qij}, if ${\bar{x}}_0=0$, the row $i=0$ of the matrix $\mathsf{A}$ is zero. As a consequence, the row $i=0$ and the column $j=0$ of the matrices $\mathsf{B}^{-1}$ and $\mathsf{B}$ have the forms
\beq
B_{0j}^{-1}=\delta_{j0},\quad B_{i0}^{-1}=-A_{i0}, \quad i\geq 1,
\eeq
\beq
B_{0j}=\delta_{j0},\quad B_{i0}=\sum_{k=1}^n B_{ik}A_{k0}, \quad i\geq 1.
\eeq
Thus, application of Eq.\ \eqref{Gij} to the pair $i$-$0$ with $i\geq 1$ yields
\beq
G_{i0}(s)=s^{-2}e^{-\sigma_{i0}s}L_{i0}(s)+\sum_{j=1}^n G_{ij}(s)A_{j0}(s).
\label{Gi0}
\eeq
Therefore, the cross
function $G_{i0}(s)$ (with $i=1,\ldots,n$), which is related to the spatial correlation
between a particle of species $i\geq 1$ and the single particle $i=0$, is
expressed in terms of the matrix $G_{ij}(s)$ of the $n$-component
mixture and the cross elements $L_{i0}(s)$ and $A_{j0}(s)$.

In principle, the nonadditivity of the $i$-$0$ interaction would be measured by the nonadditivity parameter $\Delta_{i0}$ defined by $\sigma_{i0}=\frac{1}{2}(\sigma_0+\sigma_i)(1+\Delta_{i0})$. However, the use of $\Delta_{i0}$ is not convenient in the wall limit $\sigma_0\to\infty$ that we will take at the end. Instead, we define a nonadditivity distance $\ellw_i$ by
$\sigma_{i0}=\frac{1}{2}(\sigma_0+\sigma_i)+\ellw_i$. Note that, since no $0$-$0$ interaction is present, the definition of the diameter $\sigma_0$ is somewhat arbitrary. In fact, if all $\ellw_i=\ellw$ are equal, the apparently nonadditive $i$-$0$ interaction is indistinguishable from an additive interaction with $\sigma_0\to\sigma_0+2\ellw$. Therefore, a true nonadditive $i$-$0$ interaction requires, first, that $n\geq 2$ and, second, that not all $\{\ellw_i\}$ are equal. Therefore, without loss of generality, we take $\min(\ellw_i; i=1,\ldots,n)=0$. This defines the diameter $\sigma_0$ unambiguously.

As a next step toward the wall limit, we introduce the shifted radial distribution function
\beq
\gamma_i(z)=g_{i0}(z+\sigma_0/2).
\label{gamma}
\eeq
Thus, while $r$ is the distance between the centers of the pair $i$-$0$, $z=r-\frac{1}{2}\sigma_0$ represents the distance from the center of a sphere of species $i$ to  the surface of the single sphere $j=0$. If we call $\Gamma_i(s)$ the Laplace transform of $\gamma_i(z)$, the following relationship applies:
\beq
G_{i0}(s)=e^{-\sigma_0 s/2}\left[\frac{\sigma_0}{2}\Gamma_i(s)-\Gamma_i'(s)\right],
\label{Gamma}
\eeq
where $\Gamma_i'(s)=\partial \Gamma_i(s)/\partial s$.

Finally,  we take the wall limit $\sigma_0\to\infty$. In that
case, the function $\gamma_i(z)$ becomes the ratio between
the local {number} density of particles of species $i$ at a distance
$z$ from the wall, $\rho_i(z)$, and the corresponding density in the
\emph{bulk}, {$\rho_i^{\text{b}}$. In an infinite system (as implicitly assumed in the theoretical approach), the bulk and average values coincide, i.e., $\rho_i^{\text{b}}=\rho_i(\infty)=\bar{x}_i\bar{\rho}$}.

In the wall limit $\Gamma_i'(s)$ can be neglected versus
$\sigma_0\Gamma_i(s)/2$ in Eq.\ \eqref{Gamma}, so that
\beqa
\label{wall-limit}
\Gamma_i(s)&=&2\lim_{\sigma_0\to\infty}\sigma_0^{-1}e^{\sigma_0s/2}G_{i0}(s)\nn
&=& 2e^{-(\sigma_i/2+\ellw_i)s} \frac{\widetilde{L}_i(s)}{s^2}+2\sum_{j=1}^n G_{ij}(s)\widetilde{A}_j(s),\nn
\eeqa
where in the second step we have made use of Eq.\ \eqref{Gi0} and have defined
\beq
\widetilde{L}_i(s)\equiv\lim_{\sigma_0\to\infty}\sigma_0^{-1} L_{i0}(s),
\label{Li}
\eeq
\beq
\widetilde{A}_j(s)\equiv\lim_{\sigma_0\to\infty}\sigma_0^{-1}e^{\sigma_0 s/2}A_{j0}(s).
\label{Aj}
\eeq
These two quantities are evaluated in the Appendix.

Once the Laplace transform $\Gamma_i(s)$ is well defined, let us consider the correction described by the second line of Eq.\ \eqref{54}. First, we subtract $\frac{1}{2}\sigma_0$ to the distances, so that the shell $\min(\sigma_{i0},\s_{i0})\leq r\leq \max(\sigma_{i0},\s_{i0})$ becomes $\min\left(\frac{1}{2}\sigma_{i}+\ellw_i,\widetilde{\s}_{i}\right)\leq z\leq \max\left(\frac{1}{2}\sigma_{i}+\ellw_i,\widetilde{\s}_{i}\right)$, where $\widetilde{\s}_{i}$ is the
minimum of the list of values $\sigma_{ik}-\frac{1}{2}\sigma_{k}$ ($k=1,\ldots,n$) that differ from $\frac{1}{2}\sigma_{i}+\ellw_i$. Again, $\widetilde{\s}_{i}=\frac{1}{2}\sigma_{i}+\ellw_i$ if $\sigma_{ik}-\frac{1}{2}\sigma_{k}=\frac{1}{2}\sigma_{i}+\ellw_i$ for all $k$. Finally, in the limit $\sigma_0\to \infty$, one obtains
\beqa
\gamma_{i}(z)&=&\Theta\left(z-\frac{1}{2}\sigma_{i}-\ellw_i\right)\left[\mathcal{L}^{-1}\left\{\Gamma_{i}(s)\right](r)\right.\nn
&&\left.+2\widetilde{C}_{i}\Theta(\widetilde{\s}_{i}-z)\left(\widetilde{\s}_{i}-z\right)\right\},
\label{3.1}
\eeqa
with
\beq
\widetilde{C}_{i}=
2\pi{\bar{\rho}} {\bar{x}}_{\kappa_{i}} L_{i\kappa_{i}}^\one\left(
\widetilde{L}_{\kappa_{i}}^\one-\widetilde{L}^\zero\frac{\sigma_{\kappa_{i}}+\ellw_{\kappa_{i}}}{2}\right)\left(\sigma_{\kappa_{i}}+\ellw_{\kappa_{i}}\right),
\label{3.2}
\eeq
where
{$\kappa_{i}$} is the index associated with $\widetilde{\s}_{i}$, i.e.,
$\widetilde{\s}_{i}=\sigma_{i\kappa_{i}}-\frac{1}{2}\sigma_{\kappa_{i}}$, and the quantities $\widetilde{L}^\zero$ and $\widetilde{L}^\one_i$ are defined in the Appendix.

The inverse Laplace transform in Eq.\ \eqref{3.1} can be easily performed numerically \cite{AW92}. On the other hand, the density ratio $\gamma_i$ at the shortest distance from the wall $z=\frac{1}{2}\sigma_i+\ellw_i$ can be derived analytically.
{}From Eq.\ \eqref{contact} we easily obtain
\beq
\gamma_i\left(z=\frac{1}{2}\sigma_i+\ellw_i\right)=2\widetilde{L}_i^\one+2\widetilde{C}_i\left(\widetilde{\s}_i-\frac{1}{2}\sigma_i-\ellw_i\right).
\label{contactgamma}
\eeq

The fact that the general scheme gives well defined expressions in the wall limit (${\bar{x}}_0=0$, $\sigma_0\to\infty$) is a stringent test on the internal consistency of the RFA approach. It also shows the convenience of dealing with explicit, analytical expressions from which the {subsequent limits} can be taken.

%%%%%%%%%%%%%%%%%%%%%%%%%%%%%%%%%%%%%%%%%%%%%%%%%%%%%%%%%%%%%%%%%%%%%%%%%%%%%%
\section{Monte Carlo simulations}
%%%%%%%%%%%%%%%%%%%%%%%%%%%%%%%%%%%%%%%%%%%%%%%%%%%%%%%%%%%%%%%%%%%%%%%%%%%%%%
\label{sec:mc}

We have simulated a binary mixture ($n=2$) of NAHS  through canonical $NVT$ MC simulations in a box of fixed volume and sides
$L_x$, $L_y$, and $L_z$ with $L_x=L_y$ and $L_z\gg L_x$. Periodic boundary conditions are enforced along the $x$ and $y$ directions, but two impenetrable hard walls are located at $z=0$ and $z=L_z$.   The
particles are initially placed on a simple cubic regular configuration
along the ${z}$ direction with a first crystal layer of particles of
species 1 juxtaposed to a crystal layer of particles of species 2. We
reject the $i$th particle move only in case of overlap with any other
particle, i.e., if $r_{ij}<\sigma_{ij}$ for some $j$, or with one of the walls,
i.e., if $\min(z_i, L_z-z_i)<\frac{1}{2}\sigma_i+\ellw_i$.
The system is  then
equilibrated for $10^7$ MC steps (where a MC step corresponds to
a single particle move) and the properties are generally averaged over
additional {$10^9$} MC steps {for production}. The maximum particle displacement, the same
along each direction, is determined during the first stage
of the equilibration run in such a way {as} to ensure an average acceptance ratio of 50\% at production time.
As a compromise between the condition $L_z\gg L_x$ and the computational need of not having {too high a} number of particles, we have taken $L_x=10\sigma_1$ and $L_z=30\sigma_1$ in all the simulations presented, {except a control case with $L_z=60\sigma_1$ (see below)}. The local density profiles {$\gamma_i(z)=\rho_i(z)/\rho_i^{\text{b}}$} are obtained, for each species, from histograms of
the $z$ coordinates of the
particles  in bins of width $0.01\sigma_1$.
{The bulk values $\rho_i^{\text{b}}$ are evaluated in the region of the simulation
box with $z\approx L_z/2$, where a negligible influence from the walls is expected. Due to the finite value of $L_z$, the bulk total density $\rho^{\text{b}}=\rho_1^{\text{b}}+\rho_2^{\text{b}}$ and the bulk mole fraction $x_1^{\text{b}}=\rho_1^{\text{b}}/\rho^{\text{b}}$ differ from their respective average values $\bar{\rho}$ and $\bar{x}_1$.}

%%%%%%%%%%%%%%%%%%%%%%%%%%%%%%%%%%%%%%%%%%%%%%%%%%%%%%%%%%%%%%%%%%%%%%%%%%%%%%
\section{Results}
%%%%%%%%%%%%%%%%%%%%%%%%%%%%%%%%%%%%%%%%%%%%%%%%%%%%%%%%%%%%%%%%%%%%%%%%%%%%%%
\label{sec:results}

\subsection{{Representative systems}}

In the binary case, there are five {independent} dimensionless parameters of the problem: the size ratio $\sigma_2/\sigma_1$, the particle-particle nonadditivity parameter $\Delta_{12}$, the particle-wall nonadditivity parameter $\max(\ellw_1,\ellw_2)/\sigma_1$ [remember that, by convention, $\min(\ellw_1,\ellw_2)=0$], the average mole fraction ${\bar{x}}_1$,  and the average reduced density ${\bar{\rho}}\sigma_1^3$. Here, $\sigma_1$ is chosen as the
diameter of the small spheres and henceforth it will be used to define the length unit.

In order to focus on the nonadditivity parameters, we have chosen $\sigma_2/\sigma_1=3$ for all the systems. Next, three classes of systems have been considered: (i) a nonadditive mixture in the presence of an additive wall, (ii) an additive mixture in the presence of a nonadditive wall, and (iii) a nonadditive mixture with a nonadditive wall. As representative examples of class (i) we have chosen an equimolar mixture with either positive (system A of Table \ref{tab}) or negative (system B of Table \ref{tab}) nonadditivity and a mixture with an excess of small spheres and negative nonadditivity at two densities (systems C1 and C2, respectively). As examples of class (ii), we have chosen an equimolar mixture where the wall {presents an extra repulsion to} either the large spheres (system D) or the small spheres (system E). Finally, class (iii) is represented by {system F, which is} analogous to system D, except that the mixture has a negative nonadditivity. The reduced densities ${\bar{\rho}}\sigma_1^3$ range from $\frac{1}{30}$ to $\frac{1}{5}$, {so that the total number of particles $N=\bar{\rho}L_x^2L_z$ ranges from $100$ to $600$}. {It is also convenient to measure the density} in terms of the \emph{effective} packing fraction ${\bar{\eta}}_{\text{eff}}=\frac{\pi}{6}{\bar{\rho}}\sum_{i,j}{\bar{x}}_i{\bar{x}}_j\sigma_{ij}^3$ related to van der Waals's one-fluid theory \cite{HL70}, whose values are indicated in the last column of Table \ref{tab}. {In the low-density regime, two mixtures with the same value of $\bar{\eta}_{\text{eff}}$ would have the same compressibility factor.}

\begin{table}
\caption{{Values of the nonadditivity parameters ($\Delta_{12}$, $\ellw_1$, and $\ellw_2$), the average mole fraction ($\bar{x}_1$), and the average density (${\bar{\rho}}$) for the representative systems considered in this work. In all the cases $\sigma_2/\sigma_1=3$. The table also includes the  values ($x_1^{\text{b}}$ and $\rho^{\text{b}}$) measured in the bulk region $z\approx L_z/2$ in our MC simulations with $L_z/\sigma_1=30$.}
\label{tab}}
\begin{ruledtabular}
\begin{tabular} {ccccccccc}
Label&$\Delta_{12}$&$\ellw_1/\sigma_1$&$\ellw_2/\sigma_1$&${\bar{x}_1}$&${x_1^{\text{b}}}$&${\bar{\rho}}\sigma_1^3$&${\rho^{\text{b}}\sigma_1^3}$&${\bar{\eta}}_{\text{eff}}$
\\
\hline
A&$0.2$&$0$&$0$&$0.5$&$0.469$&$\frac{1}{30}$&$0.0337$&$0.243$\\
B&$-0.2$&$0$&$0$&$0.5$&$0.503$&$\frac{1}{20}$&$0.0513$&$0.237$\\
C1&$-0.2$&$0$&$0$&$0.9$&$0.896$&$\frac{1}{10}$&$0.1025$&$0.095$\\
C2&$-0.2$&$0$&$0$&$0.9$&$0.898$&$\frac{1}{5}$&$0.2040$&$0.190$\\
D&$0$&$0$&$0.35$&$0.5$&$0.475$&$\frac{1}{30}$&$0.0345$&$0.192$\\
E&$0$&$0.35$&$0$&$0.5$&$0.511$&$\frac{1}{20}$&$0.0503$&$0.288$\\
F&$-0.2$&$0$&$0.35$&$0.5$&$0.486$&$\frac{1}{30}$&$0.0350$&$0.158$\\
\end{tabular}
\end{ruledtabular}
\end{table}

\subsection{{Bulk values}}

{The bulk values $x_1^{\text{b}}$ and $\rho^{\text{b}}$ measured in the MC simulations with $L_z=30\sigma_1$ are also included in Table \ref{tab}. In all the cases the bulk density $\rho^{\text{b}}$ is larger than the average density $\bar{\rho}$. This is due to the fact that the effective length available to the spheres of species $i$ is not $L_z$ but $L_z-\left(\sigma_i+2\ellw_i\right)$. As a consequence, the larger deviation between $\rho^{\text{b}}$ and $\bar{\rho}$ takes place for systems D ($3.5\%$) and F ($5.0\%$), i.e., those systems where the walls produce an extra repulsion ($\ellw_2/\sigma_1=0.35$) on the big spheres. This compression effect is only partially compensated by the accumulation of particles at contact with the walls. In the case of the bulk mole fraction $x_1^{\text{b}}$, the situation is less obvious. Note the identity}
\beq
{\frac{\bar{x}_1\bar{\rho}}{x_1^{\text{b}}\rho^{\text{b}}}=\frac{1}{L_z/2}\int_{\sigma_1/2+\ellw_1}^{L_z/2} dz\, \gamma_1(z).}
\label{5.1}
\eeq
{Even if the right-hand side of Eq.\ \eqref{5.1} is generally smaller than 1, the fact that $\rho^{\text{b}}>\bar{\rho}$ can give rise to $x_1^{\text{b}}<\bar{x}_1$; i.e., the bulk would be richer in big spheres than on average. This is what actually happens for systems A, C1, C2, D, and F. This effect is especially important in systems A and D since in those cases the right-hand side of Eq.\ \eqref{5.1} turns out to be larger than 1 (see Figs.\ \ref{fig:gr-aw-pna} and \ref{fig:gr-naw-pna} below for a visual confirmation). Exceptions to the property $x_1^{\text{b}}<\bar{x}_1$ are represented by systems B and E. In those cases, the right-hand side of Eq.\ \eqref{5.1} is sufficiently smaller than 1 (see Figs.\ \ref{fig:gr-aw-nna} and \ref{fig:gr-naw-nna} below) as to compensate for the ratio $\rho^{\text{b}}/\bar{\rho}>1$.}

{Now we turn our attention to the density profiles. When presenting the theoretical RFA results for each system we have used two criteria. In the first criterion, the quantities $\bar{\rho}$ and $\bar{x}_1$ appearing in the theoretical scheme described in Sec.\ \ref{sec:rfa} have been identified with the average values employed in the simulations. In the second criterion, the RFA quantities $\bar{\rho}$ and $\bar{x}_1$ have been identified with the bulk values $\rho^{\text{b}}$ and $x_1^{\text{b}}$ found in the MC simulations with $L_z=30\sigma_1$. As said before, the theoretical approach deals with formally infinite systems ($L_z\to\infty$) where the average and bulk quantities coincide. However, when making contact with simulation data corresponding to finite $L_z$ the use of either the average or the bulk values in the RFA may be important.}

%%%%%%%%%%%%%%%%%%%%%%%%%%%%%%%%%%%%%%%%%%%%%%%%%%%%%%%%%%%%%%%%%%%%%%%%%%%%%%
\subsection{Nonadditive mixture and  additive wall}
%%%%%%%%%%%%%%%%%%%%%%%%%%%%%%%%%%%%%%%%%%%%%%%%%%%%%%%%%%%%%%%%%%%%%%%%%%%%%%
\label{sec:aw}

\begin{figure}
\begin{center}
\includegraphics[width=9cm]{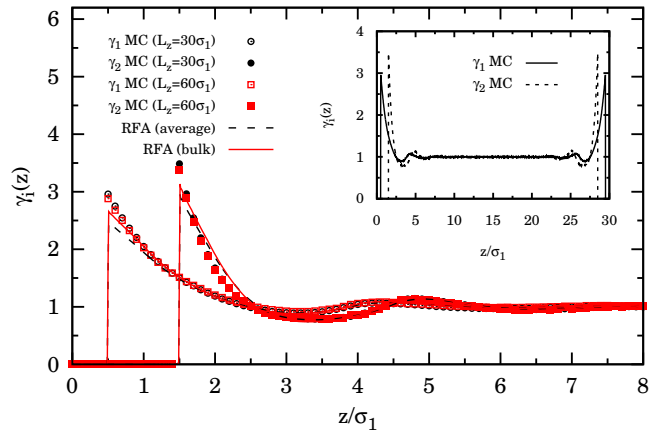}
\end{center}
\caption{{(Color online)} Local density profiles $\gamma_i(z)=\rho_i(z)/{\rho_i^{\text{b}}}$ for system A {($\sigma_2/\sigma_1=3$, $\Delta_{12}=0.2$, $\ellw_1=\ellw_2=0$, $\bar{x}_1=0.5$, $\bar{\rho}\sigma_1^3=1/30$)}.
{The lines represent the RFA theoretical predictions using the average values $\bar{x}_1$ and $\bar{\rho}$ (dashed lines) or the empirical bulk values $x_1^{\text{b}}$ and $\rho^{\text{b}}$ (solid lines). The symbols represent our MC simulations with $L_z/\sigma_1=30$ (circles) or $L_z/\sigma_1=60$ (squares). The inset shows the MC data in the whole domain $0<z<L_z$ with $L_z/\sigma_1=30$.} In the MC results, the error bars are
  within the size of the symbols used in the graph.}
\label{fig:gr-aw-pna}
\end{figure}
\begin{figure}
\begin{center}
\includegraphics[width=9cm]{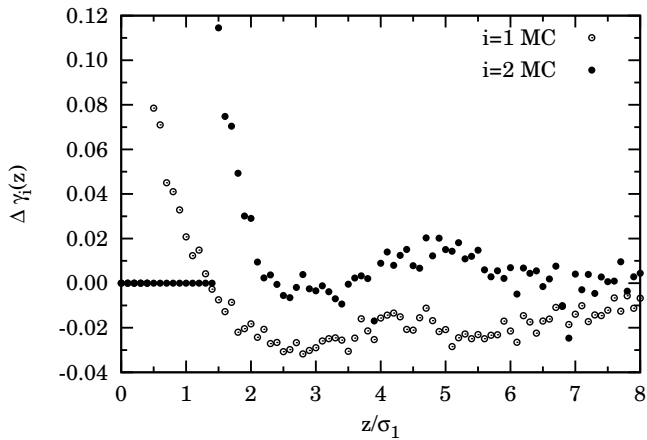}
\end{center}
\caption{{Differences $\Delta \gamma_i(z)=\left.\gamma_i(z)\right|_{30}-\left.\gamma_i(z)\right|_{60}$ between the local densities  $\gamma_i(z)=\rho_i(z)/\rho_i^{\text{b}}$ evaluated in MC simulations with $L_z/\sigma_1=30$ and those evaluated in MC simulations with $L_z/\sigma_1=60$ for system A ($\sigma_2/\sigma_1=3$, $\Delta_{12}=0.2$, $\ellw_1=\ellw_2=0$, $\bar{x}_1=0.5$, $\bar{\rho}\sigma_1^3=1/30$)}.}
\label{fig:size}
\end{figure}

Figure \ref{fig:gr-aw-pna} shows the MC and RFA results for the two (relative) density profiles $\gamma_i(z)=\rho_i(z)/{\rho_i^{\text{b}}}$ ($i=1,2$) in the case of system A (positive nonadditivity). In this system  $\widetilde{\s}_1=0.9\sigma_1>\frac{1}{2}\sigma_1$ and $\widetilde{\s}_2=1.9\sigma_1>\frac{1}{2}\sigma_2$, so that the correction term given by the second line of Eq.\ \eqref{3.1} is used {in the RFA curves}.

{The inset of Fig.\ \ref{fig:gr-aw-pna} shows the MC results for both density profiles in the whole region $0<z<L_z=30\sigma_1$. We can see that the separation between both hard walls is large enough as to identify a well defined bulk region in the center. We have chosen system A to assess the influence of finite $L_z$ by carrying out a control simulation with $L_z=60\sigma_1$. The new bulk values are $x_1^{\text{b}}=0.485$ and $\rho^{\text{b}}\sigma_1^3=0.0334$, which, as expected, are closer to the average values than in the case $L_z=30\sigma_1$  {(see Table \ref{tab})}. As seen from Fig.\ \ref{fig:gr-aw-pna}, the MC data obtained with $L_z=30\sigma_1$ and $L_z=60\sigma_1$ are hardly distinguishable, except near contact where the smaller system, having a larger bulk density, presents slightly higher values of $\gamma_i(z)$. A more detailed comparison is made in Fig.\ \ref{fig:size}, where the differences between the values of $\gamma_i(z)$ as obtained with both values of $L_z$ are shown. Figure \ref{fig:size} confirms that the smaller system ($L_z=30\sigma_1$) presents larger values for the two reduced densities near contact than the larger system ($L_z=60\sigma_1$). For {higher} separations the differences are much {less important}, but yet it is interesting to note that the smaller system tends to present larger values of $\gamma_2(z)$ but smaller values of $\gamma_1(z)$.}

{Now let us go back to Fig.\ \ref{fig:gr-aw-pna} and comment on the performance of the RFA.} We observe that the RFA underestimates the local densities at contact (i.e., at $z=\frac{1}{2}\sigma_i$). On the other hand, the decay of the local densities near the walls and the subsequent oscillations are very well captured by the theory. {It is interesting to remark that the agreement with the MC data near contact improves when the bulk values instead of the average ones are used in the theory.}

\begin{figure}
\begin{center}
\includegraphics[width=9cm]{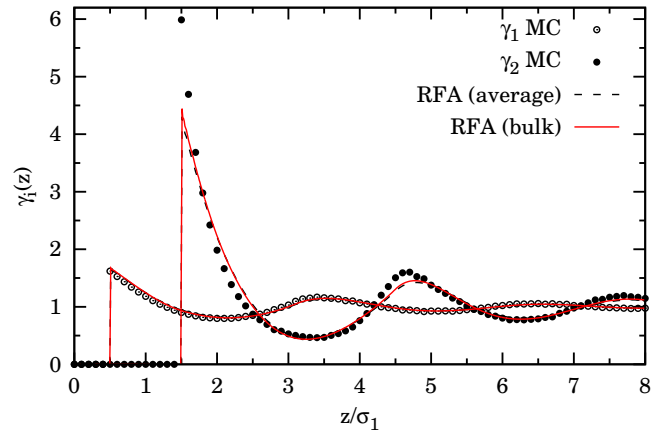}
\end{center}
\caption{{(Color online) Local density profiles $\gamma_i(z)=\rho_i(z)/\rho_i^{\text{b}}$ for system B ($\sigma_2/\sigma_1=3$, $\Delta_{12}=-0.2$, $\ellw_1=\ellw_2=0$, $\bar{x}_1=0.5$, $\bar{\rho}\sigma_1^3=1/20$).
The lines represent the RFA theoretical predictions using the average values $\bar{x}_1$ and $\bar{\rho}$ (dashed lines) or the empirical bulk values $x_1^{\text{b}}$ and $\rho^{\text{b}}$ (solid lines). The symbols represent our MC simulations with $L_z/\sigma_1=30$. In the MC results, the error bars are
  within the size of the symbols used in the graph.}}
\label{fig:gr-aw-nna}
\end{figure}

The profiles for system B (negative nonadditivity) are displayed in Fig.\ \ref{fig:gr-aw-nna}. In this case  $\widetilde{\s}_1=0.1\sigma_1$ and $\widetilde{\s}_2=1.1\sigma_1$. Since $\widetilde{\s}_i<\frac{1}{2}\sigma_i$, the
correction term in the second line of Eq.\ \eqref{3.1} vanishes.
{Comparison between Figs.\ \ref{fig:gr-aw-pna}  and \ref{fig:gr-aw-nna}  shows that, in going from system A to system B,  the local variation of the density of the big spheres is enhanced, while the local density of the small spheres becomes less structured. Here there are two competing effects at play. On the one hand, at a fixed density, the change from positive to negative nonadditivity  produces a weaker density structure near the wall, as the exact result to first order in density clearly shows. On the other hand, at a fixed nonadditivity, an increase in density  induces a higher structure. It seems that, in the transition from system A to system B, the latter effect dominates in the case of the big spheres (which are very weakly influenced by the small component) and the former effect does it in the case of the small spheres (which are strongly influenced by the presence of the large component). It is interesting to note that all these} features are very well described by the RFA, especially in the case of $\gamma_1(z)$. {The contact value of $\gamma_2$ is better estimated in system A than in system B, while the opposite happens for the contact value of $\gamma_1$. Note also that a small discrepancy is observed near the second peak of  $\gamma_2(z)$ in Fig.\ \ref{fig:gr-aw-nna}. For this system the RFA is practically insensitive to the use of the bulk values instead of the average ones.}

In systems A and B the big spheres occupy as much as 27 times more volume than the small ones, so the global properties of the mixture are dominated by species 2. A more balanced situation takes place in systems C1 and C2, where the ratio of partial packing fractions is ${\bar{x}}_2\sigma_2^3/{\bar{x}}_1\sigma_1^3=3$. In these cases the high concentration asymmetry requires a long simulation run time to reach thermal equilibrium for the big spheres.

\begin{figure}
\begin{center}
\includegraphics[width=9cm]{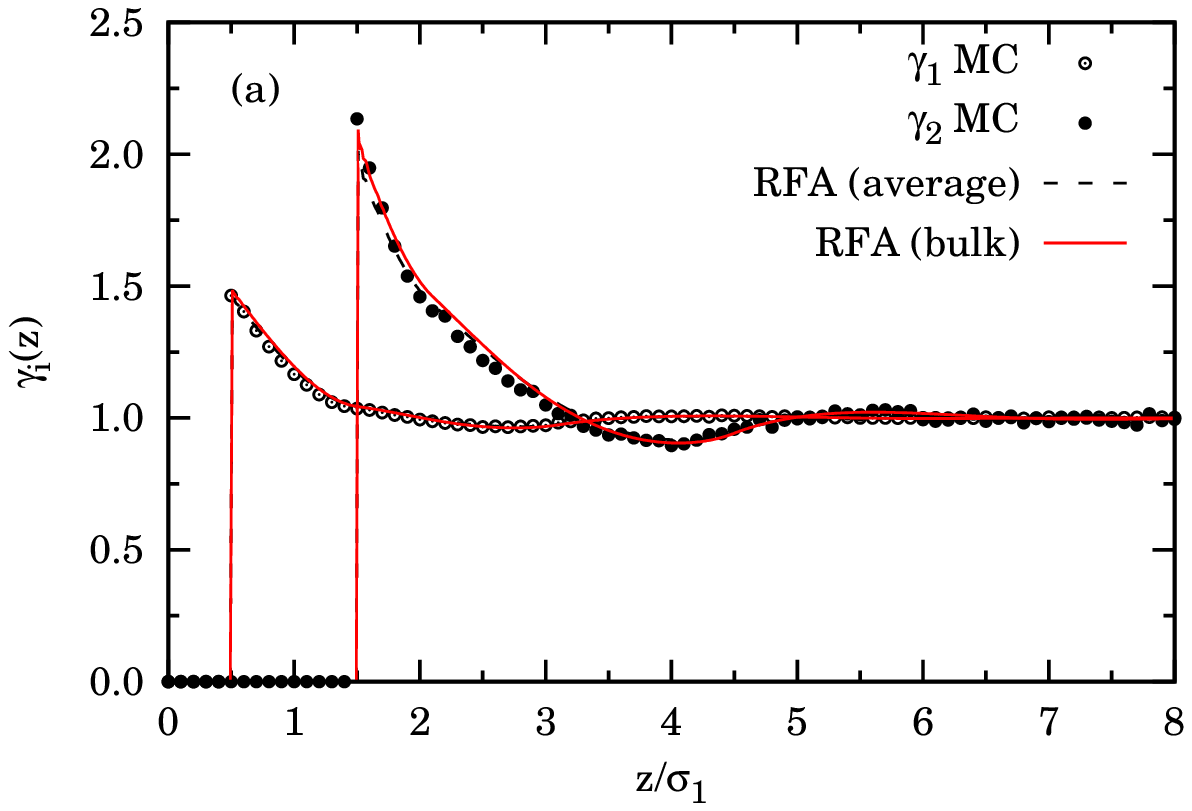}
\includegraphics[width=9cm]{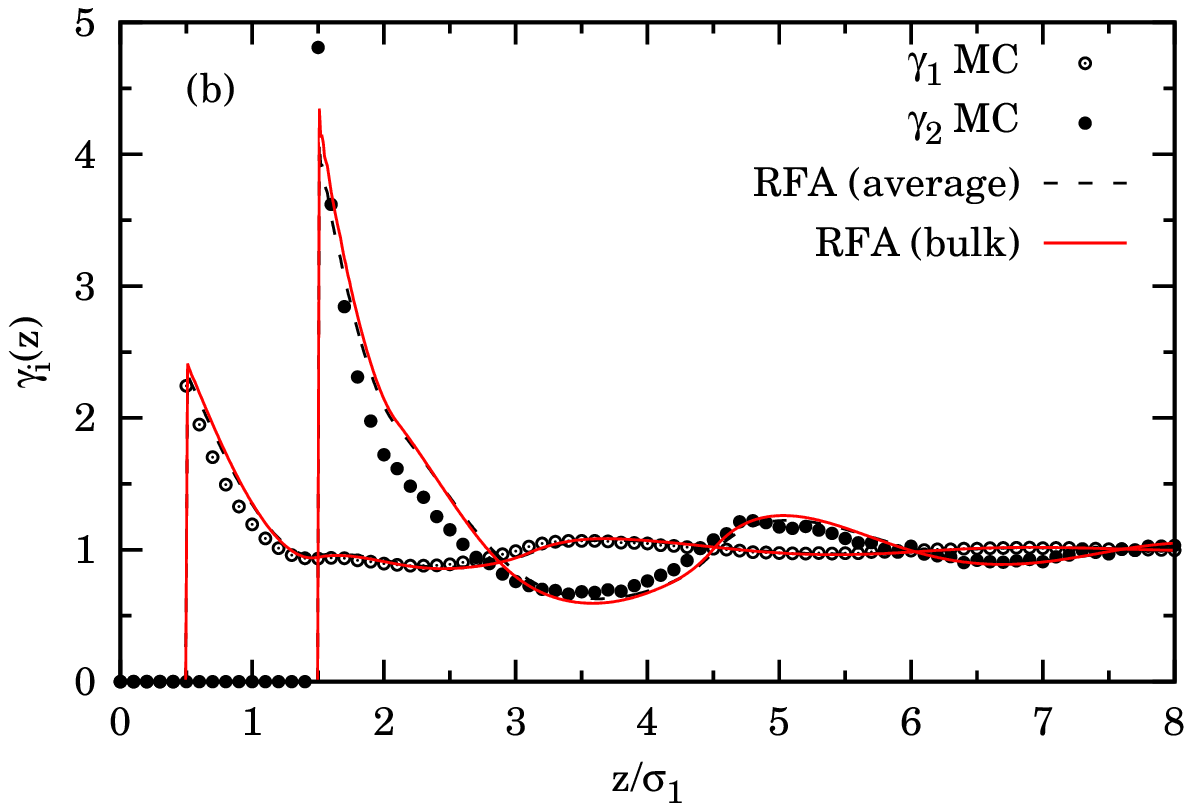}
\end{center}
\caption{{(Color online) Local density profiles $\gamma_i(z)=\rho_i(z)/\rho_i^{\text{b}}$ for (a) system C1 ($\sigma_2/\sigma_1=3$, $\Delta_{12}=-0.2$, $\ellw_1=\ellw_2=0$, $\bar{x}_1=0.9$, $\bar{\rho}\sigma_1^3=1/10$) and (b) system C2 ($\sigma_2/\sigma_1=3$, $\Delta_{12}=-0.2$, $\ellw_1=\ellw_2=0$, $\bar{x}_1=0.9$, $\bar{\rho}\sigma_1^3=1/5$).
The lines represent the RFA theoretical predictions using the average values $\bar{x}_1$ and $\bar{\rho}$ (dashed lines) or the empirical bulk values $x_1^{\text{b}}$ and $\rho^{\text{b}}$ (solid lines). The symbols represent our MC simulations with $L_z/\sigma_1=30$. In the MC results, the error bars are
  within the size of the symbols used in the graph.}}
\label{fig:asy}
\end{figure}

The results for systems C1 and C2 are shown in  Fig.\ \ref{fig:asy}. At the smaller density (system C1) the agreement between theory and simulation is almost perfect. As the density is doubled (system C2), some small deviations are visible, especially in the case of the big spheres. {Again, the RFA with the bulk values behaves near contact better than with the average values.}

%%%%%%%%%%%%%%%%%%%%%%%%%%%%%%%%%%%%%%%%%%%%%%%%%%%%%%%%%%%%%%%%%%%%%%%%%%%%%%
\subsection{Additive mixture and  nonadditive wall}
%%%%%%%%%%%%%%%%%%%%%%%%%%%%%%%%%%%%%%%%%%%%%%%%%%%%%%%%%%%%%%%%%%%%%%%%%%%%%%
\label{sec:naw}

Now we consider the cases where the mixture is additive but the wall treats differently both species. The extra repulsion affects the big spheres in system D and the small spheres in system E. In both cases $\widetilde{\s}_i\leq \frac{1}{2}\sigma_i+\ellw_i$, so that again the correction term in Eq.\ \eqref{3.1} does not apply.

\begin{figure}
\begin{center}
\includegraphics[width=9cm]{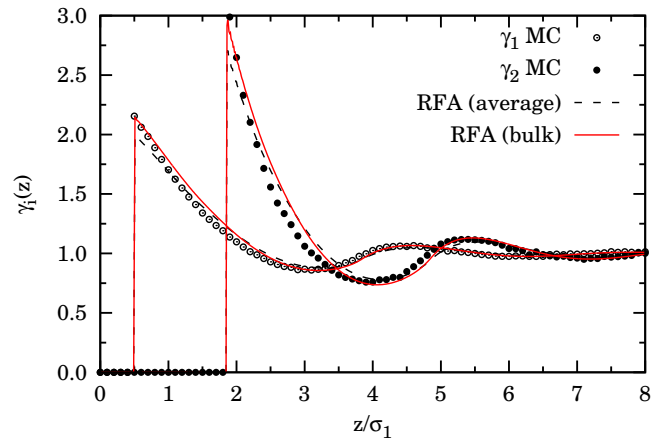}
\end{center}
\caption{{(Color online) Local density profiles $\gamma_i(z)=\rho_i(z)/\rho_i^{\text{b}}$ for  system D ($\sigma_2/\sigma_1=3$, $\Delta_{12}=0$, $\ellw_1=0$, $\ellw_2/\sigma_1=0.35$, $\bar{x}_1=0.5$, $\bar{\rho}\sigma_1^3=1/30$).
The lines represent the RFA theoretical predictions using the average values $\bar{x}_1$ and $\bar{\rho}$ (dashed lines) or the empirical bulk values $x_1^{\text{b}}$ and $\rho^{\text{b}}$ (solid lines). The symbols represent our MC simulations with $L_z/\sigma_1=30$. In the MC results, the error bars are
  within the size of the symbols used in the graph.}}
\label{fig:gr-naw-pna}
\end{figure}
\begin{figure}
\begin{center}
\includegraphics[width=9cm]{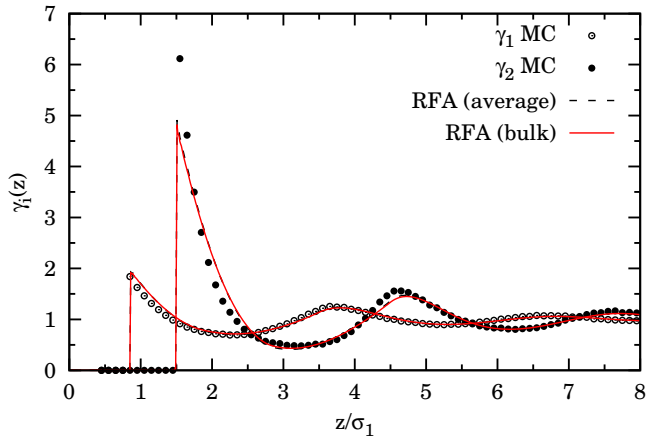}
\end{center}
\caption{{(Color online) Local density profiles $\gamma_i(z)=\rho_i(z)/\rho_i^{\text{b}}$ for  system E ($\sigma_2/\sigma_1=3$, $\Delta_{12}=0$, $\ellw_1/\sigma_1=0.35$, $\ellw_2=0$, $\bar{x}_1=0.5$, $\bar{\rho}\sigma_1^3=1/20$).
The lines represent the RFA theoretical predictions using the average values $\bar{x}_1$ and $\bar{\rho}$ (dashed lines) or the empirical bulk values $x_1^{\text{b}}$ and $\rho^{\text{b}}$ (solid lines). The symbols represent our MC simulations with $L_z/\sigma_1=30$. In the MC results, the error bars are
  within the size of the symbols used in the graph.}}
\label{fig:gr-naw-nna}
\end{figure}

The results for systems D and E are shown in Figs.\ \ref{fig:gr-naw-pna} and \ref{fig:gr-naw-nna}, respectively.
In the case of system D there is much more room for the  small spheres to sit between the wall and the big spheres than in the case of system E. As a consequence, the big spheres ``feel'' the presence of the wall more in the latter case than in the former and, thus, the contact value and the oscillations are more pronounced in system E. {These effects are enhanced by the larger density of system E relative to that of system D.} {However, $\gamma_1(z)$ near contact is higher in system D than in system E, so that the effect of wall nonadditivity compensates for the increase of density in the case of the small spheres, analogously to what happens with systems A and B (see Figs.\ \ref{fig:gr-aw-pna} and \ref{fig:gr-aw-nna}).}
All these features are correctly accounted for by the RFA, although the quantitative agreement near contact is again worse than that after the first minimum, especially  in the case of $\gamma_2(z)$. {Note also that the influence on the RFA curves of the use of the bulk versus the average values is noticeable in system D but not in system E.}

%%%%%%%%%%%%%%%%%%%%%%%%%%%%%%%%%%%%%%%%%%%%%%%%%%%%%%%%%%%%%%%%%%%%%%%%%%%%%%
\subsection{Nonadditive mixture and  nonadditive wall}
%%%%%%%%%%%%%%%%%%%%%%%%%%%%%%%%%%%%%%%%%%%%%%%%%%%%%%%%%%%%%%%%%%%%%%%%%%%%%%
\label{sec:naw2}

\begin{figure}
\begin{center}
\includegraphics[width=9cm]{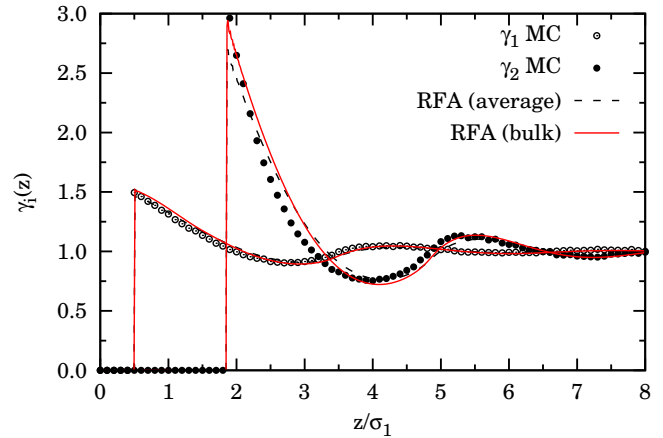}
\end{center}
\caption{{(Color online) Local density profiles $\gamma_i(z)=\rho_i(z)/\rho_i^{\text{b}}$ for  system F ($\sigma_2/\sigma_1=3$, $\Delta_{12}=-0.2$, $\ellw_1=0$, $\ellw_2/\sigma_1=0.35$, $\bar{x}_1=0.5$, $\bar{\rho}\sigma_1^3=1/30$).
The lines represent the RFA theoretical predictions using the average values $\bar{x}_1$ and $\bar{\rho}$ (dashed lines) or the empirical bulk values $x_1^{\text{b}}$ and $\rho^{\text{b}}$ (solid lines). The symbols represent our MC simulations with $L_z/\sigma_1=30$. In the MC results, the error bars are
  within the size of the symbols used in the graph.}}
\label{fig:gr-naw-pna-nna}
\end{figure}

The more general situations where both the particle-particle and the wall-particle interactions are nonadditive is, of course, richer than the preceding classes. As a simple representative system we consider the same case as in system D (wall {additionally} repelling the big spheres), except that, in addition, species 1 and 2 interact with negative nonadditivity. The resulting system F (see Table \ref{tab}) is also close to system B, except that now the wall is nonadditive and the density is smaller. As in systems B--E, the correction term in Eq.\ \eqref{3.1} is not needed.

The local densities for system F are plotted in Fig.\ \ref{fig:gr-naw-pna-nna}. Comparison with Fig.\ \ref{fig:gr-naw-pna} shows that the density profile of the big spheres is practically unaffected by the nonadditive character of the 1-2 interaction. This is not surprising taking into account that, as said before, the big spheres occupy 27 times more volume than the small ones and, therefore, the presence of the latter has little impact on the properties of the former. On the contrary, the nonadditivity has a large influence on the local density profile $\gamma_1(z)$. Since spheres of species 1 and 2 can overlap to a certain degree in system F, the big spheres partially alleviate the influence of the wall on the small spheres with respect to the case of system D. As a consequence, the local density of the small spheres is less structured in system F than in system D.
{Like in system D, the RFA performs very well in system F, especially when the bulk values are used.}

%%%%%%%%%%%%%%%%%%%%%%%%%%%%%%%%%%%%%%%%%%%%%%%%%%%%%%%%%%%%%%%%%%%%%%%%%%%%%%
\section{Conclusions}
%%%%%%%%%%%%%%%%%%%%%%%%%%%%%%%%%%%%%%%%%%%%%%%%%%%%%%%%%%%%%%%%%%%%%%%%%%%%%%
\label{sec:conclusions}

%\emph

In this work we have developed a simple analytical (in Laplace space) nonperturbative theory
for the local density profiles of a multicomponent fluid of NAHS
confined by an additive or nonadditive hard wall. The theoretical approach  is based on the specialization of the RFA technique recently proposed \cite{FS11} to the case where an extra single particle of diameter $\sigma_0$ is added to the mixture and {then} the limit of an infinite diameter $\sigma_0\to\infty$ {is  taken}.  The RFA
reduces to the exact solution of the PY approximation for zero nonadditivity, both in the particle-particle and in the particle-wall interactions, but remains  analytical even when nonadditivity prevents one from obtaining an analytical solution of the PY theory.

While the theory applies to any number of components, we have focused on a binary mixture with a size ratio $1:3$ plus a hard wall. This has allowed us to compare the theoretical results against  exact
MC simulation. Several representative scenarios have been considered (see Table \ref{tab}): a positive (system A) or negative (systems B, C1, and C2)
NAHS fluid with an additive wall, an AHS mixture with a nonadditive wall pushing either the big  (system D) or the small  (system E) spheres, and a NAHS mixture with a
nonadditive wall (system F). In all the cases, a reasonably good agreement between our theory and the MC simulations have been found for the (relative) partial local densities $\gamma_i(z)$.
The agreement is worse near contact, where the RFA underestimates the MC values, but rapidly tends to improve for larger distances, so that the initial decay of the local densities and the subsequent oscillations are rather well captured. Note that, since the RFA can be seen as a sort of continuation of the AHS PY  solution to the NAHS realm \cite{FS11}, it is not surprising that some of the features of the PY solution remain. One of those features is the underestimation of the contact values \cite{MYSH07}. Another PY feature, namely the possibility of predicting a negative first minimum at sufficiently high densities, is also inherited by the RFA.

As shown by Figs.\ {\ref{fig:gr-aw-pna} and \ref{fig:gr-aw-nna}--\ref{fig:gr-naw-pna-nna}}, the performance of the RFA is usually better in the case of the small spheres ($i=1$) than for {the} large spheres ($i=2$). This is in part due to the physical observation that the local density structure of species 1 is milder than that of species 2. Another technical reason has to do with the fact that, {while the separation between both walls  is sufficiently large for the  spheres near a wall not to be much influenced by the presence of the other wall, the unavoidable ``compression'' effect is more important for the big spheres ($L_z/\sigma_2=10$) than for the small spheres ($L_z/\sigma_1=30$). As Fig.\ \ref{fig:size} illustrates, when the separation between both walls is doubled, the effect on the density near the walls is more pronounced for the big spheres than for the small ones. Finite-size effects are also related to the small differences between the \emph{average} densities  and their \emph{bulk} values in the central region $z\approx L_z/2$. We have checked that our theoretical approach exhibits a slightly better agreement with simulations  when the empirical bulk values are used instead of the average values.}

Our theory, being a simple analytical one, can be efficiently used to
easily extract many-body approximate properties for confined fluids  under other interesting situations different from the representative ones examined in this work. For instance, extreme cases like the
Widom--Rowlinson \cite{WR70,R71,FP04}
($\sigma_1=\sigma_2=0$ with $\sigma_{12}$ finite) or the
Asakura--Oosawa \cite{AO54,AO58} ($\sigma_1=0$ and
$\Delta_{12}>0$) confined fluids can be studied. Another avenue for application of the RFA is the depletion potential between two big spheres immersed in a sea of small spheres \cite{YSH08} interacting nonadditively with them.

\appendix*
%%%%%%%%%%%%%%%%%%%%%%%%%%%%%%%%%%%%%%%%%%%%%%%%%%%%%%%%%%%%%%%%%%%%%%%%%%%%%%
\section{Evaluation of $\widetilde{L}_i(s)$ and $\widetilde{A}_j(s)$}
%%%%%%%%%%%%%%%%%%%%%%%%%%%%%%%%%%%%%%%%%%%%%%%%%%%%%%%%%%%%%%%%%%%%%%%%%%%%%%
\label{appA}
Let us recall that $\sigma_{i0}=\frac{1}{2}(\sigma_i+\sigma_0)+\ellw_i$ with $\ellw_i\geq 0$. Therefore, according to Eq.\ \eqref{bij}, $b_{i0}=\sigma_i+\ellw_i$. Thus, Eqs.\  \eqref{30a}--\eqref{32a} yield
\beq
\Lambda_0=\frac{\sigma_0}{2} {M_{2,0}}+\frac{1}{6}{M_{3,0}}+\frac{1}{2}M_{2,1},
\label{A5}
\eeq
\beq
\Psi_0=-\frac{\sigma_0}{2} {M_{2,0}}+\frac{1}{6}{M_{3,0}}-\frac{1}{2}M_{2,1},
\label{A6}
\eeq
\beqa
\Omega_0&=&-\frac{\sigma_0^2}{4}{M_{2,0}}-\frac{\sigma_0}{2} M_{2,1}-\frac{1}{4}M_{2,2},
\label{A7}
\eeqa
where
\beq
M_{p,q}\equiv \sum_{k=1}^n {\bar{x}}_k\left(\sigma_k+\ellw_k\right)^p\ellw_k^q,\quad {q\geq 0}.
\label{A3}
\eeq

Interestingly enough, the terms proportional to $\sigma_0$ and to $\sigma_0^2$ in the denominator of Eqs.\ \eqref{28a} and \eqref{29a} cancel, so that the denominator becomes
\beq
\widetilde{D}\equiv \left(1-\frac{\pi}{6}{\bar{\rho}} {M_{3,0}}\right)^2-\frac{\pi^2}{4}{\bar{\rho}}^2\left(M_{2,1}^2-
{M_{2,0}}M_{2,2}\right).
\label{A4}
\eeq

Equations \eqref{A5}--\eqref{A4} apply to any value of $\sigma_0$. {}From Eqs.\ \eqref{Lij0}--\eqref{29a} it is easy to see that both $L_{i0}^\zero$ and $L_{i0}^\one$ are linear functions of $\sigma_0$. Thus, taking the limit \eqref{Li} one gets
\beq
\widetilde{L}_i(s)=\widetilde{L}^\zero+\widetilde{L}_i^\one s
\eeq
with
\beq
\widetilde{L}^\zero=\frac{\pi{\bar{\rho}} {M_{2,0}}}{2\widetilde{D}},
\eeq
\beq
\widetilde{L}_i^\one=\frac{1}{2\widetilde{D}}\left[1-\frac{\pi}{6}{\bar{\rho}} \left({M_{3,0}}+3M_{2,1}\right)+\left(\frac{\sigma_i}{2}+\ellw_i\right)\pi {\bar{\rho}}{M_{2,0}}\right].
\eeq
Analogously, from Eqs.\ \eqref{Qij} and \eqref{Nkj}, the limit \eqref{Aj} becomes
\beq
\widetilde{A}_j(s)= \frac{2\pi{\bar{\rho}}
{\bar{x}}_j}{s^3}e^{\sigma_j s/2}\left[\widetilde{N}_{j}(s)-\widetilde{L}_{j}(s)e^{-(\sigma_{j}+\ellw_j)s}\right],
\label{Ajbis}
\eeq
where
\beqa
\widetilde{N}_{j}(s)&=&\widetilde{L}^\zero\left[1-\left(\sigma_{j}+\ellw_j\right)s+\frac{\left(\sigma_{j}+\ellw_j\right)^2s^2}{2}\right]\nn
&&+
\widetilde{L}_j^\one s \left[1-\left(\sigma_{j}+\ellw_j\right)s\right].
\eeqa

%%%%%%%%%%%%%%%%%%%%%%%%%%%%%%%%%%%%%%%%%%%%%%%%%%%%%%%%%%%%%%%%%%%%%%%%%%%%%%
\begin{acknowledgments}
{The authors are grateful to the referees for suggestions contributing to the improvement of the paper.} R.F. would like to acknowledge the use of the computational facilities
of CINECA through the ISCRA call.
A.S. acknowledges support from the Ministerio de Ciencia e Innovaci\'on (Spain) through Grant No.\ FIS2010-16587 and  the Junta de Extremadura (Spain) through Grant No.\ GR10158, partially financed by Fondo Europeo de Desarrollo Regional (FEDER) funds.
\end{acknowledgments}
%%%%%%%%%%%%%%%%%%%%%%%%%%%%%%%%%%%%%%%%%%%%%%%%%%%%%%%%%%%%%%%%%%%%%%%%%%%%%%
\bibliographystyle{apsrev}
\bibliography{D:/Dropbox/Public/bib_files/liquid}

\begin{thebibliography}{49}
\expandafter\ifx\csname natexlab\endcsname\relax\def\natexlab#1{#1}\fi
\expandafter\ifx\csname bibnamefont\endcsname\relax
  \def\bibnamefont#1{#1}\fi
\expandafter\ifx\csname bibfnamefont\endcsname\relax
  \def\bibfnamefont#1{#1}\fi
\expandafter\ifx\csname citenamefont\endcsname\relax
  \def\citenamefont#1{#1}\fi
\expandafter\ifx\csname url\endcsname\relax
  \def\url#1{\texttt{#1}}\fi
\expandafter\ifx\csname urlprefix\endcsname\relax\def\urlprefix{URL }\fi
\providecommand{\bibinfo}[2]{#2}
\providecommand{\eprint}[2][]{\url{#2}}

\bibitem[{\citenamefont{{D. Henderson, F. F. Abraham, and J. A.
  Barker}}(1976)}]{HAB76}
\bibinfo{author}{\bibnamefont{{D. Henderson, F. F. Abraham, and J. A.
  Barker}}}, \bibinfo{journal}{Mol. Phys.} \textbf{\bibinfo{volume}{31}},
  \bibinfo{pages}{1291} (\bibinfo{year}{1976}).

\bibitem[{\citenamefont{Henderson}(1978)}]{H78}
\bibinfo{author}{\bibfnamefont{D.}~\bibnamefont{Henderson}},
  \bibinfo{journal}{J. Chem. Phys.} \textbf{\bibinfo{volume}{68}},
  \bibinfo{pages}{780} (\bibinfo{year}{1978}).

\bibitem[{\citenamefont{{M. Plischke and D. Henderson}}(1984)}]{PH84}
\bibinfo{author}{\bibnamefont{{M. Plischke and D. Henderson}}},
  \bibinfo{journal}{J. Phys. Chem.} \textbf{\bibinfo{volume}{88}},
  \bibinfo{pages}{6544} (\bibinfo{year}{1984}).

\bibitem[{\citenamefont{{M. Plischke and D. Henderson}}(1985)}]{PH85}
\bibinfo{author}{\bibnamefont{{M. Plischke and D. Henderson}}},
  \bibinfo{journal}{J. Chem. Phys.} \textbf{\bibinfo{volume}{84}},
  \bibinfo{pages}{2846} (\bibinfo{year}{1985}).

\bibitem[{\citenamefont{{D. Henderson, K.-Y. Chan, and L.
  Degr\'eve}}(1994)}]{HCD94}
\bibinfo{author}{\bibnamefont{{D. Henderson, K.-Y. Chan, and L. Degr\'eve}}},
  \bibinfo{journal}{J. Chem. Phys.} \textbf{\bibinfo{volume}{101}},
  \bibinfo{pages}{6975} (\bibinfo{year}{1994}).

\bibitem[{\citenamefont{{R. Dickman, P. Attard, and V.
  Simonian}}(1997)}]{DAS97}
\bibinfo{author}{\bibnamefont{{R. Dickman, P. Attard, and V. Simonian}}},
  \bibinfo{journal}{J. Chem. Phys.} \textbf{\bibinfo{volume}{107}},
  \bibinfo{pages}{205} (\bibinfo{year}{1997}).

\bibitem[{\citenamefont{{W. Olivares-Rivas, L. Degr\'eve, D. Henderson, and J.
  Quintana}}(1997)}]{ODHQ97}
\bibinfo{author}{\bibnamefont{{W. Olivares-Rivas, L. Degr\'eve, D. Henderson,
  and J. Quintana}}}, \bibinfo{journal}{J. Chem. Phys.}
  \textbf{\bibinfo{volume}{106}}, \bibinfo{pages}{8160} (\bibinfo{year}{1997}).

\bibitem[{\citenamefont{{J. Noworyta, D. Henderson, S. Soko{\l}owski, and J.-Y.
  Chan}}(1998)}]{NHSC98}
\bibinfo{author}{\bibnamefont{{J. Noworyta, D. Henderson, S. Soko{\l}owski, and
  J.-Y. Chan}}}, \bibinfo{journal}{Mol. Phys.} \textbf{\bibinfo{volume}{95}},
  \bibinfo{pages}{415} (\bibinfo{year}{1998}).

\bibitem[{\citenamefont{{Z. Tan, U. Marini Bettolo Marconi, F. van Swol, and K.
  E. Gubbins}}(1989)}]{TMSG89}
\bibinfo{author}{\bibnamefont{{Z. Tan, U. Marini Bettolo Marconi, F. van Swol,
  and K. E. Gubbins}}}, \bibinfo{journal}{J. Chem. Phys.}
  \textbf{\bibinfo{volume}{90}}, \bibinfo{pages}{3704} (\bibinfo{year}{1989}).

\bibitem[{\citenamefont{{I. K. Snook and D. Henderson}}(1978)}]{SH78}
\bibinfo{author}{\bibnamefont{{I. K. Snook and D. Henderson}}},
  \bibinfo{journal}{J. Chem. Phys.} \textbf{\bibinfo{volume}{68}},
  \bibinfo{pages}{2134} (\bibinfo{year}{1978}).

\bibitem[{\citenamefont{{L. Degr\'eve and D. Henderson}}(1993)}]{DH93}
\bibinfo{author}{\bibnamefont{{L. Degr\'eve and D. Henderson}}},
  \bibinfo{journal}{J. Chem. Phys.} \textbf{\bibinfo{volume}{100}},
  \bibinfo{pages}{1606} (\bibinfo{year}{1993}).

\bibitem[{\citenamefont{{M. Rottereau, T. Nicolai, and J. C.
  Gimel}}(2005)}]{RNG05}
\bibinfo{author}{\bibnamefont{{M. Rottereau, T. Nicolai, and J. C. Gimel}}},
  \bibinfo{journal}{Eur. Phys. J. E} \textbf{\bibinfo{volume}{18}},
  \bibinfo{pages}{37} (\bibinfo{year}{2005}).

\bibitem[{\citenamefont{Malijevsk\'y et~al.}(2007)\citenamefont{Malijevsk\'y,
  Yuste, Santos, and {L\'opez de Haro}}}]{MYSH07}
\bibinfo{author}{\bibfnamefont{A.}~\bibnamefont{Malijevsk\'y}},
  \bibinfo{author}{\bibfnamefont{S.~B.} \bibnamefont{Yuste}},
  \bibinfo{author}{\bibfnamefont{A.}~\bibnamefont{Santos}}, \bibnamefont{and}
  \bibinfo{author}{\bibfnamefont{M.}~\bibnamefont{{L\'opez de Haro}}},
  \bibinfo{journal}{Phys. Rev. E} \textbf{\bibinfo{volume}{75}},
  \bibinfo{pages}{061201} (\bibinfo{year}{2007}).

\bibitem[{\citenamefont{Patra and Ghosh}(1997)}]{PG97}
\bibinfo{author}{\bibfnamefont{C.~N.} \bibnamefont{Patra}} \bibnamefont{and}
  \bibinfo{author}{\bibfnamefont{S.~K.} \bibnamefont{Ghosh}},
  \bibinfo{journal}{J. Chem. Phys.} \textbf{\bibinfo{volume}{106}},
  \bibinfo{pages}{2762} (\bibinfo{year}{1997}).

\bibitem[{\citenamefont{Patra}(1999)}]{P99}
\bibinfo{author}{\bibfnamefont{C.~N.} \bibnamefont{Patra}},
  \bibinfo{journal}{J. Chem. Phys.} \textbf{\bibinfo{volume}{111}},
  \bibinfo{pages}{6573} (\bibinfo{year}{1999}).

\bibitem[{\citenamefont{Roth and Dietrich}(2000)}]{RD00}
\bibinfo{author}{\bibfnamefont{R.}~\bibnamefont{Roth}} \bibnamefont{and}
  \bibinfo{author}{\bibfnamefont{S.}~\bibnamefont{Dietrich}},
  \bibinfo{journal}{Phys. Rev. E} \textbf{\bibinfo{volume}{62}},
  \bibinfo{pages}{6926} (\bibinfo{year}{2000}).

\bibitem[{\citenamefont{Zhou and Ruckenstein}(2000)}]{ZR00}
\bibinfo{author}{\bibfnamefont{S.}~\bibnamefont{Zhou}} \bibnamefont{and}
  \bibinfo{author}{\bibfnamefont{E.}~\bibnamefont{Ruckenstein}},
  \bibinfo{journal}{J. Chem. Phys.} \textbf{\bibinfo{volume}{112}},
  \bibinfo{pages}{5242} (\bibinfo{year}{2000}).

\bibitem[{\citenamefont{Zhou}(2001)}]{Z01}
\bibinfo{author}{\bibfnamefont{S.}~\bibnamefont{Zhou}}, \bibinfo{journal}{Phys.
  Rev. E} \textbf{\bibinfo{volume}{63}}, \bibinfo{pages}{061206}
  (\bibinfo{year}{2001}).

\bibitem[{\citenamefont{Choudhury and Ghosh}(2001)}]{CG01}
\bibinfo{author}{\bibfnamefont{N.}~\bibnamefont{Choudhury}} \bibnamefont{and}
  \bibinfo{author}{\bibfnamefont{S.~K.} \bibnamefont{Ghosh}},
  \bibinfo{journal}{J. Chem. Phys.} \textbf{\bibinfo{volume}{114}},
  \bibinfo{pages}{8530} (\bibinfo{year}{2001}).

\bibitem[{\citenamefont{Patra and Ghosh}(2002{\natexlab{a}})}]{PG02a}
\bibinfo{author}{\bibfnamefont{C.~N.} \bibnamefont{Patra}} \bibnamefont{and}
  \bibinfo{author}{\bibfnamefont{S.~K.} \bibnamefont{Ghosh}},
  \bibinfo{journal}{J. Chem. Phys.} \textbf{\bibinfo{volume}{116}},
  \bibinfo{pages}{8509} (\bibinfo{year}{2002}{\natexlab{a}}).

\bibitem[{\citenamefont{Patra and Ghosh}(2002{\natexlab{b}})}]{PG02b}
\bibinfo{author}{\bibfnamefont{C.~N.} \bibnamefont{Patra}} \bibnamefont{and}
  \bibinfo{author}{\bibfnamefont{S.~K.} \bibnamefont{Ghosh}},
  \bibinfo{journal}{J. Chem. Phys.} \textbf{\bibinfo{volume}{116}},
  \bibinfo{pages}{9845} (\bibinfo{year}{2002}{\natexlab{b}}).

\bibitem[{\citenamefont{Patra and Ghosh}(2002{\natexlab{c}})}]{PG02c}
\bibinfo{author}{\bibfnamefont{C.~N.} \bibnamefont{Patra}} \bibnamefont{and}
  \bibinfo{author}{\bibfnamefont{S.~K.} \bibnamefont{Ghosh}},
  \bibinfo{journal}{J. Chem. Phys.} \textbf{\bibinfo{volume}{117}},
  \bibinfo{pages}{8933} (\bibinfo{year}{2002}{\natexlab{c}}).

\bibitem[{\citenamefont{Patra and Ghosh}(2003)}]{PG03}
\bibinfo{author}{\bibfnamefont{C.~N.} \bibnamefont{Patra}} \bibnamefont{and}
  \bibinfo{author}{\bibfnamefont{S.~K.} \bibnamefont{Ghosh}},
  \bibinfo{journal}{J. Chem. Phys.} \textbf{\bibinfo{volume}{118}},
  \bibinfo{pages}{3668} (\bibinfo{year}{2003}).

\bibitem[{\citenamefont{{Y. Duda, E. Vakarin, and J. Alejandre}}(2003)}]{DVA03}
\bibinfo{author}{\bibnamefont{{Y. Duda, E. Vakarin, and J. Alejandre}}},
  \bibinfo{journal}{J. Colloid Interf. Sci.} \textbf{\bibinfo{volume}{258}},
  \bibinfo{pages}{10} (\bibinfo{year}{2003}).

\bibitem[{\citenamefont{{A. Patrykiejew, S. Soko{\l}owski, and O.
  Pizio}}(2005)}]{PSP05}
\bibinfo{author}{\bibnamefont{{A. Patrykiejew, S. Soko{\l}owski, and O.
  Pizio}}}, \bibinfo{journal}{J. Phys. Chem. B} \textbf{\bibinfo{volume}{109}},
  \bibinfo{pages}{14227} (\bibinfo{year}{2005}).

\bibitem[{\citenamefont{{F. Jim\'enez-\'Angeles, Y. Duda, G. Odriozola, and M.
  Lozada-Cassou}}(2008)}]{JDOL08}
\bibinfo{author}{\bibnamefont{{F. Jim\'enez-\'Angeles, Y. Duda, G. Odriozola,
  and M. Lozada-Cassou}}}, \bibinfo{journal}{J. Phys. Chem. C}
  \textbf{\bibinfo{volume}{112}}, \bibinfo{pages}{18028}
  (\bibinfo{year}{2008}).

\bibitem[{\citenamefont{{P. Hopkins and M. Schmidt}}(2011)}]{HS11b}
\bibinfo{author}{\bibnamefont{{P. Hopkins and M. Schmidt}}},
  \bibinfo{journal}{Phys. Rev. E} \textbf{\bibinfo{volume}{83}},
  \bibinfo{pages}{050602} (\bibinfo{year}{2011}).

\bibitem[{\citenamefont{{R. Fantoni and A. Santos}}(2011)}]{FS11}
\bibinfo{author}{\bibnamefont{{R. Fantoni and A. Santos}}},
  \bibinfo{journal}{Phys. Rev. E} \textbf{\bibinfo{volume}{84}},
  \bibinfo{pages}{041201} (\bibinfo{year}{2011}), \bibinfo{note}{{Note} that in
  Eq.\ (2.12) the hats on the partial correlation functions should be replaced
  by tildes.}

\bibitem[{\citenamefont{Yuste et~al.}(1996)\citenamefont{Yuste, {L\'opez de
  Haro}, and Santos}}]{YHS96}
\bibinfo{author}{\bibfnamefont{S.~B.} \bibnamefont{Yuste}},
  \bibinfo{author}{\bibfnamefont{M.}~\bibnamefont{{L\'opez de Haro}}},
  \bibnamefont{and} \bibinfo{author}{\bibfnamefont{A.}~\bibnamefont{Santos}},
  \bibinfo{journal}{Phys. Rev. E} \textbf{\bibinfo{volume}{53}},
  \bibinfo{pages}{4820} (\bibinfo{year}{1996}).

\bibitem[{\citenamefont{{L\'opez de Haro} et~al.}(2008)\citenamefont{{L\'opez
  de Haro}, Yuste, and Santos}}]{HYS08}
\bibinfo{author}{\bibfnamefont{M.}~\bibnamefont{{L\'opez de Haro}}},
  \bibinfo{author}{\bibfnamefont{S.~B.} \bibnamefont{Yuste}}, \bibnamefont{and}
  \bibinfo{author}{\bibfnamefont{A.}~\bibnamefont{Santos}}, in
  \emph{\bibinfo{booktitle}{{Theory and Simulation of Hard-Sphere Fluids and
  Related Systems}}}, edited by
  \bibinfo{editor}{\bibfnamefont{A.}~\bibnamefont{Mulero}}
  (\bibinfo{publisher}{Springer-Verlag}, \bibinfo{address}{Berlin},
  \bibinfo{year}{2008}), vol. \bibinfo{volume}{753} of
  \emph{\bibinfo{series}{Lectures Notes in Physics}}, pp.
  \bibinfo{pages}{183--245}.

\bibitem[{\citenamefont{Hansen and McDonald}(2006)}]{HM06}
\bibinfo{author}{\bibfnamefont{J.-P.} \bibnamefont{Hansen}} \bibnamefont{and}
  \bibinfo{author}{\bibfnamefont{I.~R.} \bibnamefont{McDonald}},
  \emph{\bibinfo{title}{{Theory of Simple Liquids}}}
  (\bibinfo{publisher}{Academic Press}, \bibinfo{address}{London},
  \bibinfo{year}{2006}).

\bibitem[{\citenamefont{{R. Fantoni and G. Pastore}}(2003)}]{FP03}
\bibinfo{author}{\bibnamefont{{R. Fantoni and G. Pastore}}},
  \bibinfo{journal}{J. Chem. Phys.} \textbf{\bibinfo{volume}{119}},
  \bibinfo{pages}{3810} (\bibinfo{year}{2003}).

\bibitem[{\citenamefont{Gonz{\'a}lez et~al.}(2011)\citenamefont{Gonz{\'a}lez,
  White, Rom{\'a}n, and Velasco}}]{GWRV11}
\bibinfo{author}{\bibfnamefont{A.}~\bibnamefont{Gonz{\'a}lez}},
  \bibinfo{author}{\bibfnamefont{J.~A.} \bibnamefont{White}},
  \bibinfo{author}{\bibfnamefont{F.~L.} \bibnamefont{Rom{\'a}n}},
  \bibnamefont{and} \bibinfo{author}{\bibfnamefont{S.}~\bibnamefont{Velasco}},
  \bibinfo{journal}{J. Chem. Phys.} \textbf{\bibinfo{volume}{135}},
  \bibinfo{pages}{154704} (\bibinfo{year}{2011}).

\bibitem[{\citenamefont{Salsburg et~al.}(1953)\citenamefont{Salsburg, Zwanzig,
  and Kirkwood}}]{SZK53}
\bibinfo{author}{\bibfnamefont{Z.~W.} \bibnamefont{Salsburg}},
  \bibinfo{author}{\bibfnamefont{R.~W.} \bibnamefont{Zwanzig}},
  \bibnamefont{and} \bibinfo{author}{\bibfnamefont{J.~G.}
  \bibnamefont{Kirkwood}}, \bibinfo{journal}{J. Chem. Phys.}
  \textbf{\bibinfo{volume}{21}}, \bibinfo{pages}{1098} (\bibinfo{year}{1953}).

\bibitem[{\citenamefont{Lebowitz and Zomick}(1971)}]{LZ71}
\bibinfo{author}{\bibfnamefont{J.~L.} \bibnamefont{Lebowitz}} \bibnamefont{and}
  \bibinfo{author}{\bibfnamefont{D.}~\bibnamefont{Zomick}},
  \bibinfo{journal}{J. Chem. Phys.} \textbf{\bibinfo{volume}{54}},
  \bibinfo{pages}{3335} (\bibinfo{year}{1971}).

\bibitem[{\citenamefont{Heying and Corti}(2004)}]{HC04}
\bibinfo{author}{\bibfnamefont{M.}~\bibnamefont{Heying}} \bibnamefont{and}
  \bibinfo{author}{\bibfnamefont{D.~S.} \bibnamefont{Corti}},
  \bibinfo{journal}{Fluid Phase Equil.} \textbf{\bibinfo{volume}{220}},
  \bibinfo{pages}{85} (\bibinfo{year}{2004}).

\bibitem[{\citenamefont{Santos}(2007)}]{S07}
\bibinfo{author}{\bibfnamefont{A.}~\bibnamefont{Santos}},
  \bibinfo{journal}{Phys. Rev. E} \textbf{\bibinfo{volume}{76}},
  \bibinfo{pages}{062201} (\bibinfo{year}{2007}).

\bibitem[{\citenamefont{Lebowitz}(1964)}]{L64}
\bibinfo{author}{\bibfnamefont{J.~L.} \bibnamefont{Lebowitz}},
  \bibinfo{journal}{Phys. Rev.} \textbf{\bibinfo{volume}{133}},
  \bibinfo{pages}{A895} (\bibinfo{year}{1964}).

\bibitem[{\citenamefont{Yuste et~al.}(1998)\citenamefont{Yuste, Santos, and
  {L\'opez de Haro}}}]{YSH98}
\bibinfo{author}{\bibfnamefont{S.~B.} \bibnamefont{Yuste}},
  \bibinfo{author}{\bibfnamefont{A.}~\bibnamefont{Santos}}, \bibnamefont{and}
  \bibinfo{author}{\bibfnamefont{M.}~\bibnamefont{{L\'opez de Haro}}},
  \bibinfo{journal}{J. Chem. Phys.} \textbf{\bibinfo{volume}{108}},
  \bibinfo{pages}{3683} (\bibinfo{year}{1998}).

\bibitem[{\citenamefont{Rohrmann and Santos}(2011{\natexlab{a}})}]{RS11}
\bibinfo{author}{\bibfnamefont{R.~D.} \bibnamefont{Rohrmann}} \bibnamefont{and}
  \bibinfo{author}{\bibfnamefont{A.}~\bibnamefont{Santos}},
  \bibinfo{journal}{Phys. Rev. E} \textbf{\bibinfo{volume}{83}},
  \bibinfo{pages}{011201} (\bibinfo{year}{2011}{\natexlab{a}}).

\bibitem[{\citenamefont{Rohrmann and Santos}(2011{\natexlab{b}})}]{RS11b}
\bibinfo{author}{\bibfnamefont{R.~D.} \bibnamefont{Rohrmann}} \bibnamefont{and}
  \bibinfo{author}{\bibfnamefont{A.}~\bibnamefont{Santos}},
  \bibinfo{journal}{Phys. Rev. E} \textbf{\bibinfo{volume}{84}},
  \bibinfo{pages}{041203} (\bibinfo{year}{2011}{\natexlab{b}}).

\bibitem[{\citenamefont{Abate and Whitt}(1992)}]{AW92}
\bibinfo{author}{\bibfnamefont{J.}~\bibnamefont{Abate}} \bibnamefont{and}
  \bibinfo{author}{\bibfnamefont{W.}~\bibnamefont{Whitt}},
  \bibinfo{journal}{Queueing Systems} \textbf{\bibinfo{volume}{10}},
  \bibinfo{pages}{5} (\bibinfo{year}{1992}).

\bibitem[{\citenamefont{Henderson and Leonard}(1970)}]{HL70}
\bibinfo{author}{\bibfnamefont{D.}~\bibnamefont{Henderson}} \bibnamefont{and}
  \bibinfo{author}{\bibfnamefont{P.~J.} \bibnamefont{Leonard}},
  \bibinfo{journal}{Proc. Natl. Acad. Sci. USA} \textbf{\bibinfo{volume}{67}},
  \bibinfo{pages}{1818} (\bibinfo{year}{1970}).

\bibitem[{\citenamefont{Widom and Rowlinson}(1970)}]{WR70}
\bibinfo{author}{\bibfnamefont{B.}~\bibnamefont{Widom}} \bibnamefont{and}
  \bibinfo{author}{\bibfnamefont{J.}~\bibnamefont{Rowlinson}},
  \bibinfo{journal}{J. Chem. Phys.} \textbf{\bibinfo{volume}{15}},
  \bibinfo{pages}{1670} (\bibinfo{year}{1970}).

\bibitem[{\citenamefont{Ruelle}(1971)}]{R71}
\bibinfo{author}{\bibfnamefont{D.}~\bibnamefont{Ruelle}},
  \bibinfo{journal}{Phys. Rev. Lett.} \textbf{\bibinfo{volume}{16}},
  \bibinfo{pages}{1040} (\bibinfo{year}{1971}).

\bibitem[{\citenamefont{Fantoni and Pastore}(2004)}]{FP04}
\bibinfo{author}{\bibfnamefont{R.}~\bibnamefont{Fantoni}} \bibnamefont{and}
  \bibinfo{author}{\bibfnamefont{G.}~\bibnamefont{Pastore}},
  \bibinfo{journal}{Physics A} \textbf{\bibinfo{volume}{332}},
  \bibinfo{pages}{349} (\bibinfo{year}{2004}), \bibinfo{note}{{Note that there
  is a misprint in Eq.\ (13), which should read
  $\bar{h}_{12}(k)=\bar{c}_{12}(k)[1-\rho_1\rho_2\bar{c}_{12}^2(k)]^{-1}$}}.

\bibitem[{\citenamefont{Asakura and Oosawa}(1954)}]{AO54}
\bibinfo{author}{\bibfnamefont{S.}~\bibnamefont{Asakura}} \bibnamefont{and}
  \bibinfo{author}{\bibfnamefont{F.}~\bibnamefont{Oosawa}},
  \bibinfo{journal}{J. Chem. Phys.} \textbf{\bibinfo{volume}{22}},
  \bibinfo{pages}{1255} (\bibinfo{year}{1954}).

\bibitem[{\citenamefont{Asakura and Oosawa}(1958)}]{AO58}
\bibinfo{author}{\bibfnamefont{S.}~\bibnamefont{Asakura}} \bibnamefont{and}
  \bibinfo{author}{\bibfnamefont{F.}~\bibnamefont{Oosawa}},
  \bibinfo{journal}{J. Polym. Sci.} \textbf{\bibinfo{volume}{33}},
  \bibinfo{pages}{183} (\bibinfo{year}{1958}).

\bibitem[{\citenamefont{Yuste et~al.}(2008)\citenamefont{Yuste, Santos, and
  {L\'opez de Haro}}}]{YSH08}
\bibinfo{author}{\bibfnamefont{S.~B.} \bibnamefont{Yuste}},
  \bibinfo{author}{\bibfnamefont{A.}~\bibnamefont{Santos}}, \bibnamefont{and}
  \bibinfo{author}{\bibfnamefont{M.}~\bibnamefont{{L\'opez de Haro}}},
  \bibinfo{journal}{J. Chem. Phys.} \textbf{\bibinfo{volume}{128}},
  \bibinfo{pages}{134507} (\bibinfo{year}{2008}).

\end{thebibliography}

%%%%%%%%%%%%%%%%%%%%%%%%%%%%%%%%%%%%%%%%%%%%%%%%%%%%%%%%%%%%%%%%%%%%%%%%%%%%%%
%%%%%%%%%%%%%%%%%%%%%%%%%%%%%%%%%%%%%%%%%%%%%%%%%%%%%%%%%%%%%%%%%%%%%%%%%%%%%%
%%%%%%%%%%%%%%%%%%%%%%%%%%%%%%%%%%%%%%%%%%%%%%%%%%%%%%%%%%%%%%%%%%%%%%%%%%%%%%
\end{document}